\documentclass{aastex}          
\usepackage{spr-astr-addons}    
\begin{document}

\title{The AGN Hubble Diagram and Its Implications for Cosmology}

\shorttitle{The AGN Hubble Diagram}
\shortauthors{Melia}

\author{F. Melia\altaffilmark{1}}
\affil{Department of Physics, the Applied Math Program, and Department of Astronomy,
The University of Arizona, Tucson, AZ 85721 \\
E-mail: fmelia@email.arizona.edu}

\altaffiltext{1}{John Woodruff Simpson Fellow.} 

\begin{abstract}
We use a recently proposed luminosity distance measure for relatively nearby
active galactic nuclei (AGNs) to test the predicted expansion of the Universe in
the $R_{\rm h}=ct$ and $\Lambda$CDM cosmologies. This comparative study
is particularly relevant to the question of whether or not the Universe
underwent a transition from decelerated to accelerated expansion, which is
believed to have occurred---on the basis of Type Ia SN studies---within the redshift
range ($0\lesssim z\lesssim 1.3$) that will eventually be sampled by these objects. We find
that the AGN Hubble Diagram constructed from currently available sources
does not support the existence of such a transition. While the scatter
in the AGN data is still too large for any firm conclusions to be drawn, the results
reported here nonetheless somewhat strengthen similar results of comparative
analyses using other types of source. We show that the Akaike, Kullback, and 
Bayes Information Criteria all consistently yield a likelihood of $\sim 84-96\%$ 
that $R_{\rm h}=ct$ is closer to the ``true" cosmology than $\Lambda$CDM is,
though neither model adequately accounts for the data, suggesting an
unnaccounted-for source of scatter.
\end{abstract}

\keywords{cosmological parameters; cosmology: observations;
cosmology: redshift; cosmology: theory; active galactic nuclei; gravitation}

\section{Introduction}
A proposal was made recently to infer accurate luminosity distances to Active Galactic Nuclei
(AGNs) using the tight relationship (established via reverberation mapping) between the
luminosity of their central engine and the radius of the broad-line region (BLR)
\citep{Watson2011}. If feasible, this technique would open up the possibility of examining the
cosmological expansion out to a redshift $z\sim 2-3$ using a class of objects other
than the already well known and studied Type Ia supernovae \citep{Riess1998,Perlmutter1999}.

Finding reliable distance measures beyond the reach ($z\sim 2$)  of current tools is difficult,
but several methods have been proposed in the past few years. We recently added some
support to the idea of using gamma ray burst sources (GRBs) to construct a Hubble Diagram
(HD) out to redshifts $z\sim 5-6$ \citep{Wei2013}, using correlations among certain spectral
and lightcurve features as luminosity indicators. Using the most up-to-date GRB sample
appropriate for this work, we showed that the GRB HD produces fits useful in delimiting
the possible expansion scenarios in this redshift range, though $\sim 20\%$ of the events
lie at least $2\sigma$ away from the best-fit curves, suggesting that either some
contamination by non-standard GRB luminosities is unavoidable, or that the errors
and intrinsic scatter are still being underestimated. This class of sources will no doubt
become increasingly important as the precision of their measured properties continues
to improve, but there is clearly still a need to search for other possibilities.

In another study, closely related to the subject of this paper, we also proposed the
use of high-$z$ quasars to construct an HD at redshifts $z\gtrsim 6-7$ \citep{Melia2014b}.
The use of high-$z$ quasars as standard candles has recently
been made possible by the recognition that a single observation of the quasar's
spectrum can yield both its optical/UV luminosity---and therefore the distance
of line-emitting gas from the central ionizing source---and the width of BLR lines,
such as Mg II---which facilitates a measurement of the velocity of the line-emitting
gas. Together, these data can, in principle, provide an accurate determination
of the black hole's mass. And since it is becoming more and more evident that
quasars at $z\gtrsim 6$ are accreting at close to their Eddington limit
\citep{Willott2010,DeRosa2011}, it may be possible to base the high-$z$ quasar
HD on the assumption that the luminosity function at these high redshifts is
well constrained.\footnote{As we shall see shortly, this approach is quite different
from that suggested for nearby AGNs, even though both make use of our
knowledge concerning the BLR. The high-$z$ quasar technique is, by necessity,
statistical in nature, whereas the nearby AGN method relies on the measurement
of fluxes and time lags in individual sources.} Of course, to use this method
reliably, one needs to have sufficient redshift coverage. The discovery
of quasar ULAS J1120+0641 at $z=7.085$ \citep{Mortlock2011} has extended
the range of these sources sufficiently for us to begin using this approach 
in model comparisons.

Since their discovery in the early 1960's, many attempts have been made
to use AGNs as standard candles
\citep{Baldwin1977,Collier1999,Elvis2002,Marziani2003}. None of these were
very successful, but the aforementioned improvements in our understanding
of the BLR have dramatically changed this situation. In \S2 of this paper,
we will describe the method suggested by Watson et al. \citep{Watson2011}
to construct
the nearby AGN HD, and then apply it to test the predictions of several
cosmological models in \S3. One of our primary goals will be to compare
the $R_{\rm h}=ct$ Universe directly with $\Lambda$CDM in the very
important redshift range $0\lesssim z\lesssim 2$, where the best evidence for a
transition from cosmic deceleration to acceleration is claimed to have
been found. We will discuss the consequences of our results in \S4.

\section{A Distance Measure Using AGNs}
Reverberation mapping \citep{Blandford1982} relies on high-quality
spectrophotometric monitoring of an AGN over an extended period of
time (in many cases lasting several years). BLR lines are 
produced via photoionization in the hot accretion disk surrounding 
the black hole, which produces a variable continuum flux. These 
variations are echoed by changes in the flux of the broad emission lines
after a light-crossing time. This technique probes regions only 
$\sim0.01$ pc in extent at the centers of arbitrarily distant galaxies. 
As of today, reverberation mapping has yielded black-hole masses 
for over 50 AGNs \citep{Peterson2004,Bentz2009a,Bentz2013}.

One expects that $R\propto \sqrt{L}$, where $R$ is the BLR size,
set by the depth to which the gas can be photoinoized by the
central continuum \citep{Kaspi2000,Kaspi2005,Bentz2009b}.
At the same time, simple light-travel time arguments suggest that
$R\sim \tau c$, where $\tau$ is the lag time between variations
in the continuum and the response (or echo) measured with
the broad lines (typically H$\beta$ or C IV). Thus, the observable
quantity $\tau/\sqrt{F}$, where $F$ is the measured AGN
continuum flux, should be proportional to the luminosity distance
to the source, i.e.,
\begin{equation}
d_L\propto {\tau\over\sqrt{F}}\;.
\end{equation}
Both $\tau$ and $F$ are quantities that can be observed directly, 
independently of the background expansion, when the appropriate (measured)
cosmological redshift is taken into account for the purpose of
making rest-frame measurements. The luminosity distance measured in
this way is therefore completely independent of any cosmological model.

Recent improvements in the measurement of $\tau$ and $F$ have
led to a confirmation that the radius-luminosity relationship
follows the simple law implied by Equation~(1) across four
orders of magnitude in $L$ \citep{Bentz2009a,Zu2011}.
Chief among these was the successful removal
of the contaminating effects of the host galaxy, making
measurements of the lag time more precisely by re-observing
AGNs with poorly sampled light curves, and filling in the
low-luminosity end of the sample.

Our sample of 35 observed $\tau/\sqrt{F}$ values is taken from
\cite{Watson2011}, who compiled all the available
lags for the H$\beta$ line and rest-frame 5100 $\AA$ continuum
fluxes (e.g., from Bentz et al. 2009a; Denney et al. 2010). 
These have been corrected for Galactic extinction (see also
Schlegel et al. 1998; Schlafly et al. 2010), though internal extinction
corrections are available for only a few of the 35 sources in this
sample to be applied uniformly (more on this below).

The current sample of AGNs, assembled
from all available lags in the H$\beta$-line and rest-frame 5100
\AA~continuum fluxes, exhibits a tight radius-luminosity relationship indicating
that the ionization parameter and the gas density are both close to
constant across all 35 objects. This is not surprising in view of the
the locally optimally emitting cloud model \citep{Baldwin1995}. However,
the fact that the density has the same value in the BLR for all sources
and luminosities is not yet understood. But as long as the variation 
of this gas density is small, the observational uncertainties should 
dominate the scatter.

These data are shown in Figure~1, together with the best fit curves
from several cosmological models, which we will discuss in the next
section. Several of the galaxies in this plot are identified for
specific reasons. For example, the current position of NGC 7469 
is based on the updated measurement of the lag in \cite{Zu2011},
rather than from the original observation, which indicated a
significantly discrepant lag.

NGC~3227 and NGC~4051 are highlighted because these are the only
sources with direct distance estimates. However, the Tully-Fisher distance
to NGC~4015 is the less accurate of the two, so the $\tau/\sqrt{F}$ distance
relation was calibrated to the luminosity distance of galaxy NGC~3227
(see also Tonry et al. 2001). Note, however, that the uncertainty in this
calibration is relatively large. Eventually, Cepheid-derived distances
may provide a better absolute calibration.

The source NGC~5548 demonstrates the benefit to be gained from
repeated reverberation measurements, which substantially refines
the distance to any of these objects. The observational uncertainty
for this AGN is 0.05 dex (0.13 mag) after about a dozen such observations;
it is typically $\sim0.14$ dex (0.35 mag) for sources with a single
measurement. Some flux variation in the continuum over a
measurable time $\tau$ is necessary in order to infer the time
delay in the signal reaching the BLR, where the change is
echoed in the lines. But large flux variations over a time that
would affect their location $R$ would also be known within this $\tau$,
and these are not observed. Indeed, for those systems that have been observed
repeatedly, very little intrinsic variation has been seen in $\tau/\sqrt{F}$,
suggesting that flux variability contributes much less than measurement
uncertainties to the overall scatter.

Finally, a likely source of scatter is due to extinction associated with the AGN
and its host galaxy. To illustrate how significant this effect can be, Figure~1
also highlights the position of NGC~3516, which currently lies more than
1 $\sigma$ away from the best-fit curves. However, an application of
the recently measured extinction correction \citep{Denney2010} would
shift it to a position very close to these curves. But since very 
few extinction corrections are known, we are not including them for the first
analysis of this sample. It has been estimated \citep{Watson2011} that
the overall scatter may be reduced by as much as $\sim 0.08$ dex
(0.2 mag) with the accurate correction of all of the internal extinctions.
Unfortunately, only a handful of the AGNs in the sample used here
have sufficient data for the internal AGN and host-galaxy extinction to be
estimated at the present time. And in most of those cases (with the
exception of NGC~3516), the discrepancy between estimates made in
a single object are as large as the extinction correction itself (see,
e.g., Cackett et al. 2007; Bentz et al. 2009a; Denney et al. 2010).
Nonetheless, along with other the observational uncertainties,
this possible correction does contribute to the overall scatter 
of the current data about the best-fit model curves (see figure~1). 
To gauge the impact of using more representative errors on the model 
comparison, we will therefore also carry out a best-fit analysis at 
the end of \S~3 using errors that incorporate such additional 
uncertainties not currently displayed in figure~1.

\begin{figure}[h]
\includegraphics[width=\columnwidth]{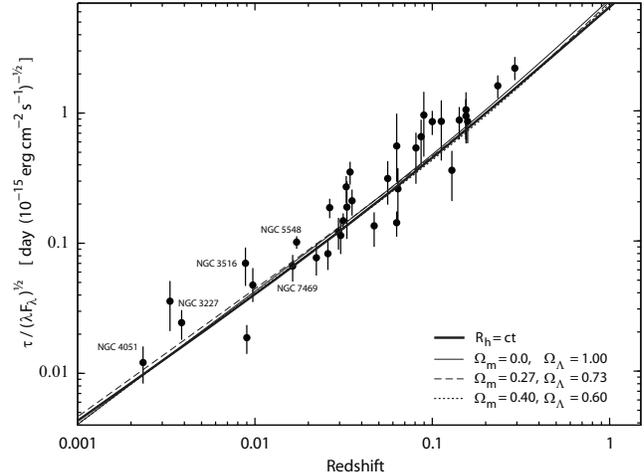}
\caption{Hubble Diagram constructed from the AGN sample in \cite{Watson2011}.
The position of NGC 7469 is based on Zu et al.'s (2011)
re-measurement of the time lags in this source using the SPEAR method. The vertical
axis shows the luminosity distance indicator $\tau/\sqrt{F}$ (see Equation~1), versus
redshift for all the AGNs with H$\beta$ measurements. Also shown are the best fit
curves for the $R_{\rm h}=ct$ Universe ($\chi^2_{\rm dof}=2.85$ for 34 degrees
of freedom), and the optimized $\Lambda$CDM model ({\it thin, solid} curve;
$\chi^2_{\rm dof}=2.97$ for 32 degrees of freedom), and two other variations
of the standard model (both with $\chi^2_{\rm dof}=3.01$ for 32 degrees of
freedom). The best fit $\Lambda$CDM model has $w_\Lambda=-1$. For the sake of
comparison, the other two variations of the standard model also have $w_\Lambda
=-1$. In addition, the $\Lambda$CDM model with $\Omega_m=0.27$ has $H_0=
74.9^{+5.3}_{-5.4}$ km s$^{-1}$ Mpc$^{-1}$ (2 model parameters and
1-$\sigma$ errors calculated as shown in Figures~2 and 3), while the model with
$\Omega_m=0.40$ has $H_0=75.0^{+4.8}_{-5.6}$ km s$^{-1}$ Mpc$^{-1}$.}
\end{figure}

\section{Theoretical Fits to the AGN Hubble Diagram}
Depending on how one chooses to characterize the dark energy and its
equation-of-state $p_\Lambda=w_\Lambda\rho_\Lambda$, $\Lambda$CDM can have
as many as 7 free parameters, including the Hubble constant $H_0$, the matter
energy density $\Omega_m\equiv \rho_m/\rho_c$ normalized to today's critical
density $\rho_c\equiv (3c^2/8\pi G)H_0^2$, the similarly defined dark energy
density $\Omega_\Lambda$, and $\Omega_k$, representing the spatial
curvature of the Universe---appearing as a term proportional to the
spatial curvature constant $k$ in the Friedmann equation. In this
paper, we will take the minimalist approach and consider only the
most essential parameters needed to fit the AGN data.  For this purpose,
we will take guidance from other observations (such as those with WMAP
and {\it Planck}), which indicate that $k=0$ (i.e., that the Universe is
spatially flat). In other words, we will treat $k$ as a prior and not
include it in the optimization procedure, which means that $\Omega_m+
\Omega_\Lambda=1$ in the redshift range of interest. As such, the
$\Lambda$CDM model we use here for comparison with the $R_{\rm h}=ct$
Universe is characterized by three essential parameters: $H_0$,
$\Omega_m$ and $w_\Lambda$, with the additional restriction that
the Universe has no phantom energy, i.e., that $w_\Lambda\ge -1$.

\begin{figure}[h]
\includegraphics[width=\columnwidth]{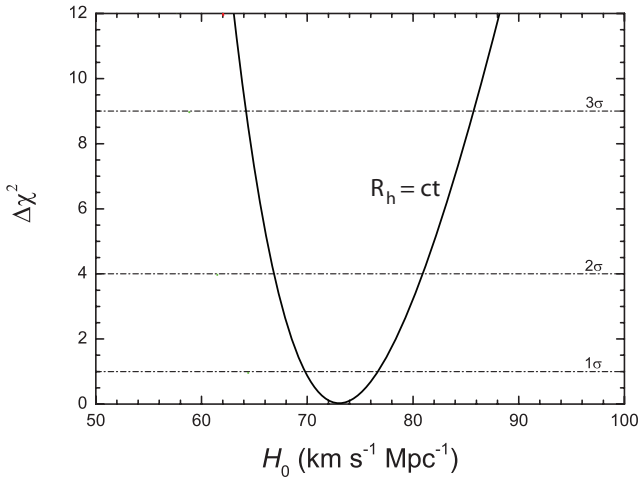}
\caption{Constraints on the Hubble constant, $H_0$, for the $R_{\rm h}=ct$
Universe, based on fits to the data shown in Figure~1. The optimized Hubble
constant has a value  $73.2^{+3.7}_{-3.5}$ km s$^{-1}$ Mpc$^{-1}$
($68.3\%$ level of confidence).}
\end{figure}

In fitting the AGN data in Figure~1, we will compare the predictions of
$\Lambda$CDM with those of the $R_{\rm h}=ct$ Universe. Both models
are Friedmann-Robertson-Walker cosmologies, but $R_{\rm h}=ct$
and $\Lambda$CDM handle $\rho$ differently. The theoretical basis
for the former is rather straightforward \citep{Melia2007,MeliaShevchuk2012} and
stems directly from the fact that the condition $R_{\rm h}=ct$,
equating the Universe's gravitational horizon (equal to the Hubble radius)
to the distance light could have traveled since the big bang, is required
by the simultaneous application of both the Cosmological principle and
Weyl's postulate \citep{Weyl1923}. This constraint forces the expansion
factor $a(t)\propto t$, which requires a total equation-of-state
$p=-\rho/3$, in terms of the total pressure $p$ and density $\rho$.
Whereas $\Lambda$CDM guesses the constituents of $\rho$ (matter,
radiation, dark energy) and their individual equations-of-state, from
which the dynamics ensues, the expansion history in $R_{\rm h}=ct$
is known precisely at all cosmic times $t$. And the various
constituents must partition themselves in such a way as to
always preserve the overall condition $p=-\rho/3$.

The observables in $R_{\rm h}=ct$ take on simple
analytic forms. For example, the luminosity distance is
given by the elegant expression\footnote{The Milne Universe \citep{Milne1933}
is sometimes confused with $R_{\rm h}=ct$, but in fact its observables are
quite different---and have already been refuted by the observations. Unlike
the $R_{\rm h}=ct$ Universe, in which the spatial curvature constant is
$k=0$, the Milne universe is empty and has $k=-1$. As a result, the luminosity
distance in Milne is $d_L^{\rm Milne}=R_{\rm h }(t_0)(1+z)
\sinh[\ln(1+z)]$, which is not at all consistent with the data
\citep{MeliaShevchuk2012}.}
\begin{equation}
D_{L}^{R_{\rm h}=ct}=\frac{c}{H_{0}}(1+z)\ln(1+z)
\end{equation}
which, unlike $\Lambda$CDM, has only one free parameter---the Hubble constant $H_0$.
The factor $c/H_0$ is in fact the gravitational horizon $R_{\rm h}(t_0)$ at the
present time, so we may also write the luminosity distance as
\begin{equation}
D_{L}^{R_{\rm h}=ct}=R_{\rm h}(t_0)(1+z)\ln(1+z)\;.
\end{equation}
By comparison, the luminosity distance in $\Lambda$CDM is given as
\begin{eqnarray}
D_{L}^{\Lambda {\rm CDM}}(z)&=&{c\over H_{0}}{(1+z)\over
\sqrt{\mid\Omega_{k}\mid}}\; {\rm sinn}\left\{\mid\Omega_{k}\mid^{1/2}
\times\right.\nonumber\\
&\null&\hskip-0.5in\left.\int_{0}^{z}{dz\over\sqrt{\Omega_m(1+z)^3+\Omega_\Lambda(1+z)^{3(1+w)}}}\right\}\;,
\end{eqnarray}
where $c$ is the speed of light. (Note that the fractional density $\Omega_r$ due to
radiation is insignificant compared to the other components, and is ignored in this expression.)
The function sinn is $\sinh$ when
$\Omega_{k}>0$ and $\sin$ when $\Omega_{k}<0$. Since we take the Universe
to be flat with $\Omega_{k}=0$, Equation (4) simplifies to the form $(1+z)c/H_{0}$
times the integral.

In Figure~1, the best fit models calculated from Equation~(3),
in the case of $R_{\rm h}=ct$, and Equation~(4), for $\Lambda$CDM,
are shown as solid curves. The Hubble constant in the case of
the former has the value $73.2^{+3.7}_{-3.5}$ km s$^{-1}$ Mpc$^{-1}$, with a
corresponding reduced $\chi^2_{\rm dof}=2.85$ (and 34 degrees of
freedom). The quality of the fit is shown as a function of $H_0$ in Figure~2,
along with the 1, 2 and 3 $\sigma$ levels of confidence.  This value is
consistent with previous measurements of the Hubble constant, e.g.,
$73.8\pm2.4$ km s$^{-1}$ Mpc$^{-1}$ reported in \cite{Riess2011},
though only marginally consistent with the Planck 2013 measurement of
$67.3\pm1.2$ km s$^{-1}$ Mpc$^{-1}$ \citep{Ade2013} . We emphasize
that this is the only free parameter available to the $R_{\rm h}=ct$ Universe.
With the conditions and constraints
described above, the best-fit $\Lambda$CDM model has the parameter
values $H_0=75.0$ km s$^{-1}$ Mpc$^{-1}$, $\Omega_m=0$, and
$w_\Lambda=-1$, with a corresponding $\chi^2_{\rm dof}=2.97$
(and 32 degrees of freedom). We note, however, that very similar
fits, characterized by comparable $\chi^2_{\rm dof}$'s, may be
obtained for $\Lambda$CDM using a wide range of parameter values,
as indicated in Figure~3, which shows the 1, 2, and 3-$\sigma$
confidence regions for a two-dimensional optimization in $\Lambda$CDM,
with the adoption of prior values for $w_{\rm de}$ and $\Omega$.
To illustrate this point, we also show in Figure~1 two other
$\Lambda$CDM models, both with $\chi^2_{\rm dof}=3.01$, and the
parameters indicated in the figure caption. In particular, the
model with $\Omega_m=0.27$ comes very close to the WMAP-concordance
model \citep{Komatsu2011}, giving confidence that the use
of AGNs to construct a Hubble diagram is meaningful.

We have also attempted to fit the data in Figure~1 using a variation
of $\Lambda$CDM without assuming flatness. For the Planck
2013 parameter values $H_0=67.3$ km s$^{-1}$ Mpc$^{-1}$
and $\Omega_{\rm m}=0.315$, the best fit corresponds to
$\Omega=1.24^{+0.19}_{-0.16}$ (1-$\sigma$ errors) and
$w_{\rm de}=-1$ (this value can range anywhere from $0$ to
$-1$ within 1-$\sigma$), with a reduced $\chi^2_{\rm dof}=2.91$
(33 degrees of freedom). The quality of this fit is comparable
to the others shown in Figure~1, so here too the AGN data
are not yet precise enough to rule out a negative spatial
curvature, though they do appear to disfavour a closed
universe.

The $\chi^2$ values in the case of $\Lambda$CDM are very similar
to those reported in \cite{Watson2011}, who discussed the
possible reasons for such large $\chi^2$'s and the likely dominant
contributions to the scatter in the AGN data shown in Figure~1.
The statistical quality of the measurements can be improved with
an increased number of observations per source, the acquisition
of more reliable lags, and better extinction estimates, all of which
could decrease the scatter in the AGN HD substantially. The claim
is that within a few years of observing, the total scatter could
be reduced to levels comparable to those of current Type Ia SN
samples \citep{Kessler2009,Conley2011}.

We can see quantitatively how significant the current uncertainties
are by redoing the analysis described above using larger errors representing
these sources of scatter. In their paper, Watson et al. (2011) considered
four different contributions, arising from observational uncertainties, 
extinction effects, bad lags, and others. They estimate an overall mean 
square scatter $\Delta m\approx 0.50$ in magnitude (see their Table 1 
for more details). The uncertainty in the measured value of $D_L$
arising from this $\Delta m$ may be added in quadrature to the errors 
shown in figure~1. Re-optimizing the $R_{\rm h}=ct$ and $\Lambda$CDM fits,
we find for the former that the reduced $\chi^2_{\rm dof}$ is now $1.258$ 
(for 34 degrees of freedom) with an inferred Hubble constant of $76.9^{+4.7}_{-5.5}$
km s$^{-1}$ Mpc$^{-1}$. The best-fit $\Lambda$CDM model has the
parameter values $H_0=78.9$ km s$^{-1}$ Mpc$^{-1}$, $\Omega_{\rm m}=0$,
and $w_\Lambda=-1$, with a corresponding $\chi^2_{\rm dof}=1.317$
(32 degrees of freedom). The quality of the $\chi^2$-fitting is
clearly better, though the optimized parameters have changed only
slightly. There may still be an additional intrinsic dispersion
that prevents $\chi^2_{\rm dof}$ from approaching unity. The 
anticipated future improvements in the precision of these
measurements should help to distinguish between competing
cosmological models compared to what can be done now.

\begin{figure}[h]
\includegraphics[width=\columnwidth]{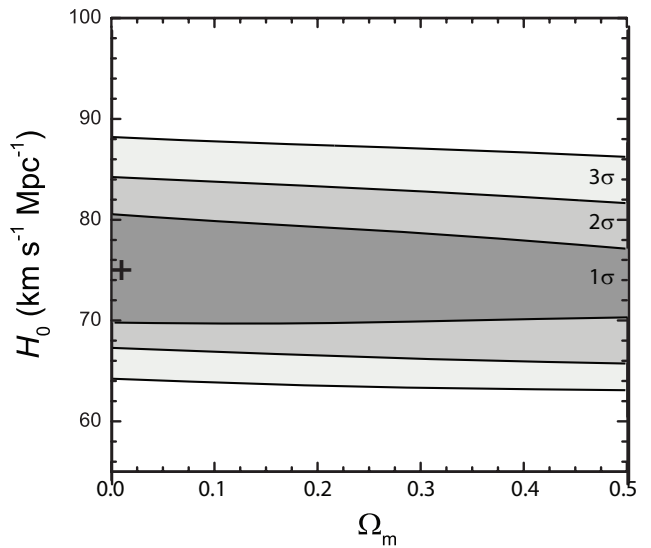}
\caption{One-, two-, and three-$\sigma$ confidence regions for a
two-dimensional optimization of the parameters $H_0$ and $\Omega_{\rm m}$
in $\Lambda$CDM, with a fixed value $w_{\rm de}=-1$ and zero
spatial curvature ($\Omega=1$). The cross indicates the parameter values
corresponding to the best-fit. As discussed in the text, $\Omega_{\rm m}$
is not yet well constrained by the AGN sample shown in Figure~1.}
\end{figure}

\section{Discussion}
The comparison between $R_{\rm h}=ct$ and $\Lambda$CDM
emerging from Figure~1 suggests that the projected improvements
in the AGN HD are important for several reasons. Unless these
results change qualitatively when the sample size and quality
improve over the coming years, $R_{\rm h}=ct$ appears
to be a better fit than $\Lambda$CDM to the AGN data.

In recent years, we have carried out this kind of comparative analysis between
the two cosmologies using a diverse range of observational data, at both low and high
redshifts. Since the formulation of measurable quantities, such as the luminosity
distance in Equations~(3) and (4), is different in $R_{\rm h}=ct$
and $\Lambda$CDM, the process of selecting the most likely correct model must also
take into account the number of free parameters. The likelikhood of either cosmology
being closer to the ``truth" may be determined from the model selection criteria
described in \cite{MeliaMaier2013}. The Akaike Information Criterion (AIC)
\citep{Liddle2004,Liddle2007,Tan2012} prefers models with few parameters to
those with many, unless the latter provide a substantially better fit to the data.
This avoids the possibility that by using a greater number of parameters, one
may simply be fitting the noise.

Information criteria were invented specifially to provide a statistical basis for
preferring one model over another when their numbers of parameters are different.
The fundamental problem is that the introduction of extra parameters will allow
an improved fit to the dataset, regardless of whether or not those new
parameters are actually relevant. A simple comparison of the maximum
likelihood of different models will therefore always favor the model with
more parameters. The information criteria were designed to compensate
for this by penalizing models that have more parameters, thereby
offsetting any improvement in the maximum likelihood allowed by the
additional parameters.

But there are different ways of implementing this idea. The two most commonly
used approaches are the AIC, which comes from minimizing the Kullback-Leibler
information entropy \citep{Takeuchi2000} (which measures the difference between
the true distribution and the model distribution), and the Bayes Information
Criterion (BIC, defined below), which
uses the posterior odds of one model against another presuming that the models
are equally favored prior to the data fitting \citep{Schwarz1978}.

Having at least two criteria helps because none of them are ideal in all circumstances.
For example, extensive Monte Carlo testing has indicated that the AIC tends to favor
models that have more parameters than the true model \citep{Harvey1993,Kass1995}.
By contrast, the BIC ever more harshly penalizes over-parameterized models as the
number of data points increases. For large datasets, BIC should therefore be preferred,
though the AIC remains useful since it gives an upper limit to the number of parameters
that ought to be included.

The AIC is defined by the expression ${\rm AIC} = \chi^2 + 2\,k$,
where $k$ is the number of free parameters. Among two models fitted to the data,
the one with the least resulting AIC is assessed as the one more likely to be ``true.''
The unnormalized confidence that model $i$ is true is the Akaike weight
$\exp(-{\rm AIC}_i/2)$.  Informally, model $i$ has likelihood
\begin{equation}
\label{eq:lastAIC}
{\cal L}_i=\frac{\exp(-{\rm AIC}_i/2)}
{\exp(-{\rm AIC}_1/2)+\exp(-{\rm AIC}_2/2)}
\end{equation}
of being closer to the correct model.  The difference ${\rm AIC}_2\nobreak-{\rm
  AIC}_1$ therefore determines the extent to which model $i$ is favored
over the other.

The Kullback Information Criterion (KIC), ${\rm KIC} = \chi^2 + 3\,k$,
is lesser known, and takes into account the fact that the PDF's of the various 
competing models may not be symmetric \citep{Cavanaugh1999}. The strength of the
evidence in KIC is similar to that for AIC, and the likelihood is calculated using
the same Equation~(5), though with AIC$_i$ replaced with KIC$_i$.
The Bayes Information Criterion (BIC) is the best known of the
three, representing an asymptotic ($N\to\infty$) approximation to the outcome of a
conventional Bayesian inference procedure for deciding between models
\citep{Schwarz1978}. This criterion is defined by ${\rm BIC} = \chi^2 + (\ln N)\,k$,
and clearly suppresses overfitting very strongly if the number of data points
$N$~is large.

We can now proceed to estimate the likelihood of either $R_{\rm h}=ct$ 
or $\Lambda$CDM being closer to the correct cosmology, based on fits to the
AGN HD in the previous section. We will do this for the re-optimization 
we carried out using the larger errors representing the scatter in the 
data (see bottom of \S~3), since the corresponding dispersions largely
mitigate the general scatter and appear to better represent the overall 
uncertainty in the current measurements. From the AIC, we infer that the
likelihood of $R_{\rm h}=ct$ being the correct cosmology is 
$\sim 84.4_{-7.8}^{+10.1}\%$ compared to $\sim 15.6_{-10.1}^{+7.8}\%$ for 
$\Lambda$CDM. The KIC results in a somewhat stronger indication, with a 
likelihood of $\sim 93.6_{-4.7}^{+4.3}\%$ that $R_{\rm h}=ct$ is correct, 
compared to $\sim 6.4_{-4.3}^{+3.7}\%$ for $\Lambda$CDM. The BIC produces 
the strongest result, mainly because the number of data points is quite 
large. According to this criterion, $R_{\rm h}=ct$ is $\sim 96.2_{-2.3}^{+2.6}\%$ 
versus only $\sim 3.8_{-2.6}^{+2.3}\%$ more likely to be correct than
$\Lambda$CDM. The errors on these likelihoods are calculated from the
$1$-$\sigma$ change in $\chi^2$, i.e., $\Delta\chi^2=1$ in the case
of $R_{\rm h}=ct$ (1 parameter) and $\Delta\chi^2=2.3$ for $\Lambda$CDM
(2 parameters).

It is important to emphasize the fact that these results are fully
consistent with, and strongly reinforce, previous results using
other observations. For example, based on the cosmic chronometer
data, we found that the likelihood of $R_{\rm h}=ct$ being closer
than $\Lambda$CDM to the correct cosmology is $\sim 82-91\%$
versus $\sim 9-18\%$ \citep{MeliaMaier2013}; from the GRB Hubble
Diagram, we found that $R_{\rm h}=ct$ is more likely than $\Lambda$CDM
to be correct with a likelihood of $\sim 85-96\%$ versus $\sim 4-15\%$
\citep{Wei2013}; from the high-$z$ quasar Hubble Diagram,
we inferrred a relative likelihood of $\sim 70\%$ versus $\sim 30\%$
in favor of $R_{\rm h}=ct$ \cite{Melia2014a}; and from the cluster gas
mass fraction data, we found this ratio to be $\sim 95\%$ over
$\sim 5\%$ \citep{Melia2013}.

\section{Conclusions}
The work reported in this paper has the
potential of impacting one of the most important results of Type Ia SN
studies---that the Universe is currently experiencing a phase of
accelerated expansion---because the two data sets will cover essentially
the same redshift range ($0\lesssim z\lesssim 2$). In this paper, we have
demonstrated that the current AGN data already favor the $R_{\rm h}=ct$
Universe over $\Lambda$CDM. However, the expansion rate in this
Universe is constant; the Universe experienced no early deceleration,
nor a current acceleration. The Type Ia SN claim of a transition from
one to the other is therefore not confirmed by the AGN HD.

Which data should we trust more? Without question, the AGN
observations are not yet precise enough to challenge Type Ia SNe.
The scatter seen in the AGN Hubble Diagram (Figure~1) is far too large.
However, we demonstrated that when some of the errors possibly
responsible for this scatter are included in the $\chi^2$-minimization
procedure, the information criteria skew the relative likelihoods even
more towards $R_{\rm h}=ct$. It is therefore likely that when the refinements 
and improvements discussed above are implemented, the results described 
in this paper will reinforce the results of other comparative tests carried 
out thus far between $R_{\rm h}=ct$ and $\Lambda$CDM, which clearly
favor the former over the latter. The BIC, in particular, consistently shows
that the likelihood of $R_{\rm h}=ct$ being correct is an overwhelming
$\gtrsim 95\%$ compared to only $\lesssim 5\%$ for $\Lambda$CDM.

What then are we to make of the Type Ia SN results? The truth is
that unlike the cosmic chronometers and the AGN HD, the Type Ia
SN data cannot be reduced independently of the pre-assumed
cosmological model. The four so-called ``nuisance" parameters
used to match the SN characteristics to a standard candle must
be optimized along with the free parameters of the adopted
background cosmology which, up until now, has always been
$\Lambda$CDM. The inherent weakness of this approach, and
the negative impact of trying to use these model-dependent data
for comparative studies, have been described in greater detail
in \cite{Melia2012} and \cite{Wei2015}. The bottom line is that if it turns out that the
Universe is expanding at a constant rate, then trying to fit the
Type Ia SN data with $\Lambda$CDM is equivalent to attempting
a cubic polynomial fit to a straight line: it is impossible to fit the
linear dependence perfectly, and the apparent transition from
an early deceleration to a current acceleration may simply be
the negative consequence of this imperfect polynomial approximation.
In view of all the evidence now available from other data, the Type Ia
SN data must be recalibrated for $R_{\rm h }=ct$ in order to complete
a proper comparative test between this cosmology and $\Lambda$CDM.
We ourselves have recently completed such a test using the Supernova
Legacy Survey Sample \citep{Guy2010}, and have
reported the results in \cite{Wei2015}. The direct one-on-one
comparison between $R_{\rm h}=ct$ and $\Lambda$CDM using this sample shows
that both models fit the data with the same $\chi^2_{\rm dof}$. This in
itself is quite important because it suggests that the optimization of the
nuisance parameters, along with the model, make the data somewhat
compliant to the assumed cosmology. The data reduced with one
model are not the same as those associated with another; yet
each set is fit equally well by its corresponding model. The conclusion
that the Universe is accelerating is therefore heavily model
dependent. But more than this, as we have found in this paper, the
AIC, KIC, and (especially) the BIC strongly favour $R_{\rm h}=ct$
because it accounts for the measurements with only one free
parameter, whereas $\Lambda$CDM has several, depending on
how one models the dark-energy equation of state.

Having said this, the importance of a high-precision AGN HD extends
well beyond the range of redshifts accessible with these sources and the
Type Ia SNe. The discourse concerning whether or not the Universe
went through a transition from deceleration to acceleration also bears
considerably on our interpretation of the cosmic microwave background
(CMB) fluctuations, and their correspondence to physics in the early
Universe. $\Lambda$CDM has had considerable success accounting for the
CMB power spectrum, certainly on scales less than a few degrees
\citep{Bennett2011}. Some of the strongest evidence in favor of the
standard model comes from our analysis and interpretation of the
radiation produced near the surface of last scattering \citep{Ade2013}.
So one must be wary about too easily discarding a model that has
enjoyed this type of success over many years.

But there are good reasons for also continuing to probe the standard
model, not only because we still lack a complete understanding of
what happened in the first few seconds following the big bang,
but also because in spite of its success, the improving precision
of our cosmological measurements points to areas of tension
between its predictions and the data. In the CMB, for example,
there are still unresolved questions concerning the emergence
of possible anomalies that may conflict with the excellent fits
to the power spectrum on small scales. For example, the
angular correlation function of the CMB not only requires
significant cosmic variance to bring theory in line with
observations but, more tellingly, reveals an absence of
any correlation at angles greater than about 60 degrees
\citep{Copi2010,Melia2014a}. This is potentially quite serious
because such an absence of angular correlation would be
inconsistent with inflationary theory, the bedrock of modern
cosmology. Without inflation, however, a standard model
with early deceleration and subsequent acceleration would
not be able to explain the general uniformity of the CMB---the
so-called horizon problem. This is why in concert with improving
Type Ia SNe observations, the construction of a precision AGN
HD may help answer the question of whether the Universe did
in fact go through a transition from deceleration to acceleration,
which bears strongly on the scientific justification for taking
inflation seriously.

\acknowledgments
I am grateful to the anonymous referee for very helpful comments that
have led to an improvement in the manuscript. I am also grateful to 
Amherst College for its support through a John Woodruff Simpson
Lectureship, and to Purple Mountain Observatory in Nanjing, China, for its hospitality
while part of this work was being carried out. This work was partially supported by
grant 2012T1J0011 from The Chinese Academy of Sciences Visiting Professorships for
Senior International Scientists, and grant GDJ20120491013 from the Chinese State
Administration of Foreign Experts Affairs.

\end{document}